\documentclass[showpacs,epsfig,aps]{revtex4}
\usepackage{epsfig}
\usepackage{amssymb}
\usepackage{amsmath}
\usepackage{graphicx}
\usepackage{lscape}
\usepackage{booktabs}
\usepackage{subfig}
\usepackage{array}

\begin{document}

\title{The specific charged hadron multiplicity in $e^-$+p and $e^-$+D
semi-inclusive deep-inelastic scattering in the PYTHIA and PACIAE models}
\author{Yu-Liang Yan$^1$ \footnote{yanyl@ciae.ac.cn}, Xing-Long Li$^1$,
        Xiao-Mei Li$^1$, Dai-Mei Zhou$^2$, Yun Cheng$^2$, Bao-Guo Dong$^1$,
        Xu Cai$^2$, and Ben-Hao Sa$^{1,2}$ \footnote{sabh@ciae.ac.cn}}

\affiliation{$^1$ China Institute of Atomic Energy, P. O. Box 275 (10),
              Beijing, 102413 China. \\
             $^2$ Key Laboratory of Quark and Lepton Physics (MOE) and
              Institute of Particle Physics, Central China Normal University,
              Wuhan 430079, China.}

\begin{abstract}
We employed the PYTHIA 6.4 model and the extended parton and hadron cascade
model PACIAE 2.2 to comparatively investigate the DIS normalized specific
charged hadron multiplicity in the 27.6 GeV electron semi-inclusive
deep-inelastic scattering off proton and deuteron. The PYTHIA and PACIAE
results calculated with default model parameters not well and fairly well
reproduce the corresponding HERMES data, respectively. In addition, we
have discussed the effects of the differences between the PYTHIA and
PACIAE models.
\end{abstract}
\maketitle

\section {Introduction}
Since the eighties of last century the lepton inclusive and semi-inclusive
deep inelastic scattering (DIS and SIDIS) off nuclear target have become
one of the most active frontiers between the nuclear and particle physics.
They have greatly contributed to the partonic structure of hadron \cite{jixd},
the parametrization of parton distribution function (PDF) \cite{plac}, and
the nuclear medium effect on PDF (EMC effect) \cite{rith}. They also play
important role in the hadronization of initial partonic state, the
space-time evolution of the fragmentation process, and the extraction
of polarization-averaged fragmentation function (FF) \cite{lead,herm1}.

The multiplicity data of specific charged hadron ($\pi^+$, $\pi^-$, $K^+$,
$K^-$) in the unpolarized SIDIS are crucial for distinguishing the
quark fragmentation function of $D_q^h$ from the antiquark one
of $D_{\bar q}^h$. Thus those multiplicity data are important for a reliable extraction of the FF. Recently, the HERMES collaboration has measured the charged pions and kaons multiplicity in the 27.6 GeV electron SIDIS off proton
and deuteron \cite{herm2}. Meanwhile, they have compared their DIS
normalized data to the HERMES Lund Monte Carlo (HLMC) simulations with
thirteen model parameters tuned to the multiplicity as a function of
$z$, $p_T$ (hadron transverse momentum), and $\eta$ (hadron
pseudorapidity) of the $\pi^-$, $K^-$, and $\bar p$ \cite{herm2,hill}.
HLMC is a combination of the DIS event generator Lepto \cite{inge}
(based on JETSET 7.4 and PYTHIA 5.7 \cite{sjos1}), the detector
simulation program (based on GEANT \cite{geant}), and the HERMES
reconstruction program \cite{hill}.
\begin{center}
\begin{figure}[]
\includegraphics[width=0.25\textwidth]{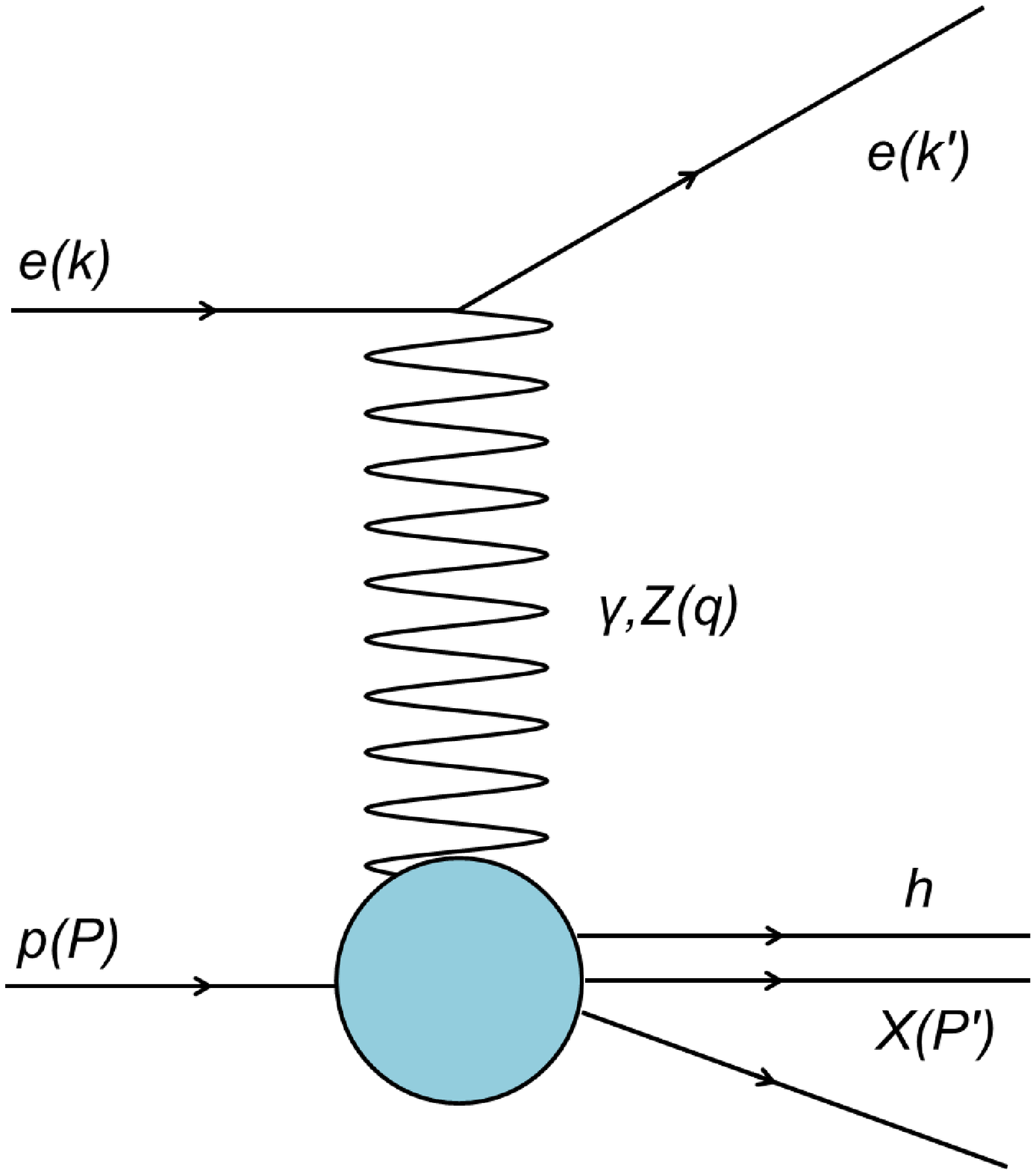} \hspace{2.5cm}
\includegraphics[width=0.25\textwidth]{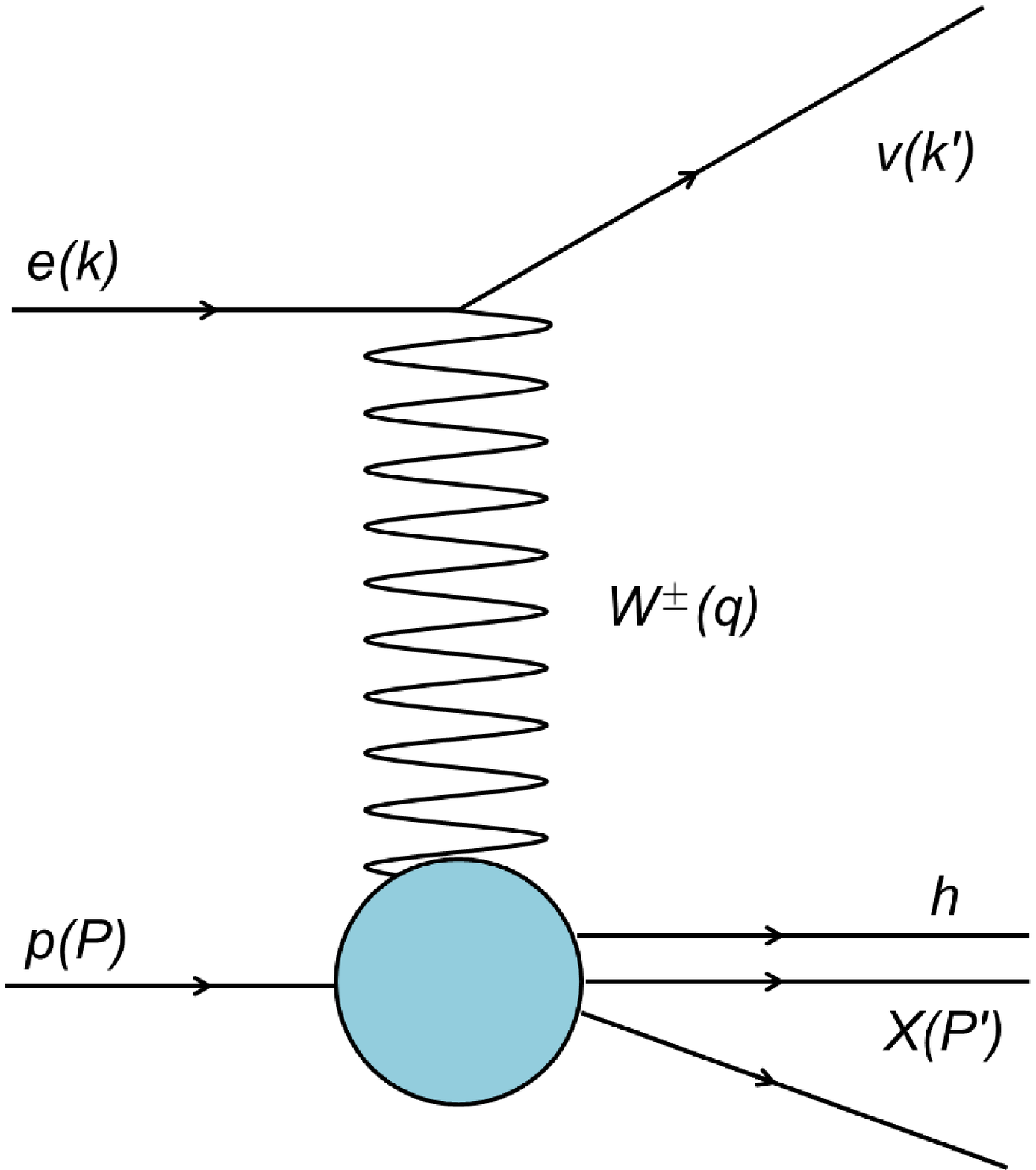}
\caption{(color online) Leading order (Born approximation) Feynman diagram
of the neutral current (NC, left panel) and charged current (CC, right panel)
$e^-$+p DIS.}
\label{nccc}
\end{figure}
\end{center}

In this paper the PYTHIA 6.4 \cite{sjos} model and on which based model of
PACIAE 2.2 \cite{sa3} (simplified as PYTHIA and PACIAE, respectively, later)
were employed to calculate the DIS normalized $\pi^+$, $\pi^-$, $K^+$, and
$K^-$ multiplicities in the 27.6 GeV electron SIDIS off the proton and
deuteron targets. The PACIAE 2.2 model is a new issue of the PACIAE model
updated presently from PACIAE 2.1 \cite{sa2} in order to cover the lepton
DIS (SIDIS) off the nuclear target. The DIS normalized specific charged
hadron multiplicity as a function of $z$ in the $e^-$+p and $e^-$+D SIDIS
at 27.6 GeV beam energy calculated by PYTHIA and PACIAE with default
model parameters is not well and fairly well consistent with the HERMES
data \cite{herm2}, respectively. The default PYTHIA results in $e^-$+D
SIDIS is calculated as the average of the default PYTHIA results in
$e^-$+p and $e^-$+n (neutron) SIDIS at the same beam energy.
\section {Models}
The PACIAE model \cite{sa2} is based on PYTHIA \cite{sjos}. However,
the PYTHIA model is for high energy elementary collision ($e^+e^-$,
lepton-hadron, and hadron-hadron ($hh$) collisions) but PACIAE is also
for lepton-nuclear and nuclear-nuclear collisions.

In the PYTHIA model, a $hh$ ($pp$) collision for instance, is described
in term of parton-parton collisions. The hard parton-parton collision
is dealt by the LO-pQCD parton-parton cross section with the modification
of parton distribution function in a hadron. The soft parton-parton
collision, a non-perturbative process, is considered empirically. The
initial- and final-state QCD radiations as well as the multiparton
interactions are taken into account. So the consequence of a $hh$
collision is a partonic multijet state composed of the diquarks
(anti-diquarks), quarks (antiquarks), and the gluons, besides a few
hadronic remnants. All of the above duarks (antiquarks), diquarks
(anti-diquarks), and gluons are constructed into strings which are
then hadronized by Lund string fragmentation regime, thus a final
hadronic state is obtained for a $hh$ ($pp$) collision eventually.

Correspondingly, in the PACIAE model a $hh$ ($pp$) collision is also
described by PYTHIA as mentioned in last pragraph. However, in the
PACIAE model, the above constructed strings are not hadronized
immediately, but are split into quarks (antiquarks), diquarks
(anti-diquarks), and gluons by force. And the diquarks (anti-diquarks)
are split further into quarks (anti-quarks). Thus one obtains a partonic
initial state composed of quarks, ant-quarks, and gluons for a $hh$
($pp$) collision. The partons then take part in the partonic rescattering
. After partonic rescattering, the partons are constructed into
strings. The strings are then hadronized by Lund string fragmentation
model. After hadronization, the produced hadrons proceed hadronic
rescattering. And the final hadronic state is reached for a hh (pp)
collision eventually.

In PACIAE model a nucleus-nucleus collision is described as follows:
The nucleons in a colliding nucleus are first randomly distributed
according to the Woods-Saxon distribution in the spatial phase space.
The participant nucleons, resulted from Glauber model calculation, are
required to be inside the overlap zone, formed when two colliding nuclei
path through each other at a given impact parameter. The spectator
nucleons are required to be outside the overlap zone but inside the
nucleus-nucleus collision system. If the beam direction is set
on $z$ axis, then $p_x=p_y=0$ and $p_z=p_{beam}$ are set for nucleons
in the projectile nucleus for both the fixed target and collider.
$p_x=p_y=p_z=0$ and $p_x=p_y=0$ as well as $p_z=-p_{beam}$ are set for
nucleons in the target nucleus for both the fixed target and collider,
respectively. We then decompose a nucleus-nucleus collision into
nucleon-nucleon ($NN$) collision pairs according to the nucleon
straight-line trajectories and the $NN$ total cross section. Each $NN$
collision is dealt by PYTHIA with string fragmentation switched-off
and diquarks (anti-diquarks) broken into quarks (anti-quarks) as
mentioned in last paragraph. A partonic initial state, composed of
the quarks, antiquarks, gluons, and a few hadronic remnants, is obtained
for a nucleus-nucleus collision when $NN$ collision pairs were exhausted.
\begin{center}
\begin{figure}[]
\includegraphics[width=0.8\textwidth]{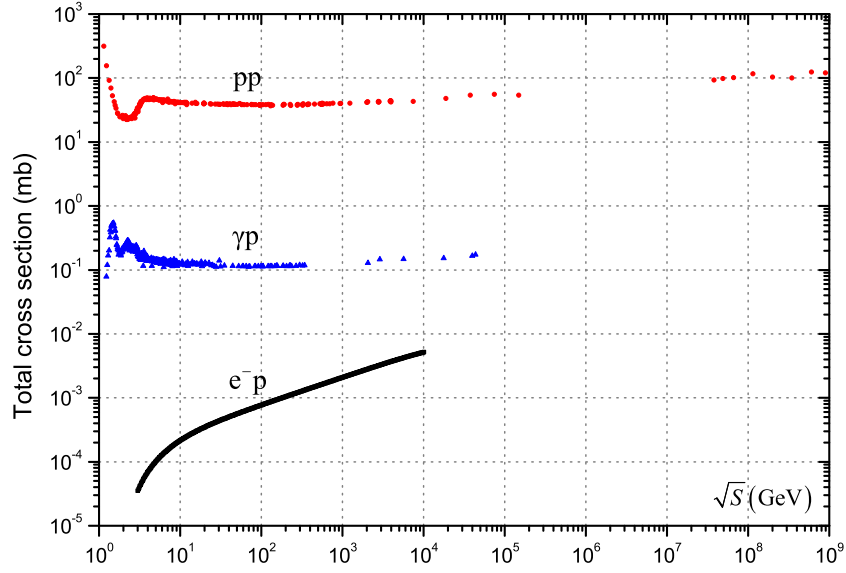}
\caption{(color online) The total cross section of pp and $\gamma$p
collisions as well as leading order $e^-$p DIS.}
\label{tcros}
\end{figure}
\end{center}

\begin{widetext}
\begin{center}
\begin{figure}[]
\vspace{1cm}
\includegraphics[width=0.5\textwidth]{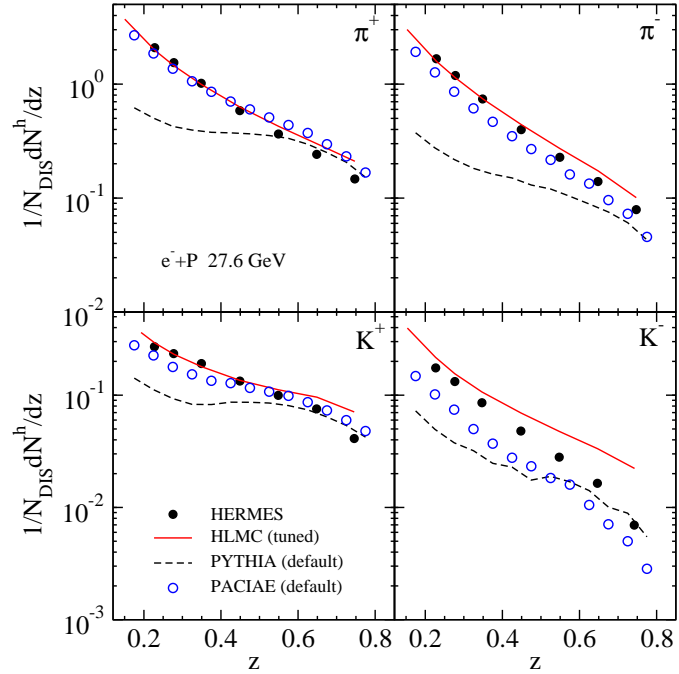}
\caption{(color online) Multiplicity of DIS normalized specific charged
hadron as a function of $z$ in the $e^-+$p SIDIS at 27.6 GeV beam
energy.}
\label{ep45}
\end{figure}
\end{center}
\end{widetext}

This partonic initial stage is followed by a parton evolution stage.
In this stage, the parton rescattering is performed by the Monte Carlo
method with $2\rightarrow2$ LO-pQCD parton-parton cross sections \cite{comb}.
The hadronization stage follows the parton evolution stage. The Lund string
fragmentation model and a phenomenological coalescence model are provided for
the hadronization. However, the string fragmentation model is selected in the
present calculations. Then the rescattering among produced hadrons is dealt
with usual two body collision model \cite{sa2}. In this hadronic evolution
stage, only the rescatterings among $\pi$, $K$, $\rho (\omega)$, $\phi$, $p$,
$n$, $\Delta$, $\Lambda$, $\Sigma$, $\Xi$, $\Omega$, and their antiparticles
are considered for simplicity.

The p+A (A+p) collisions are simulated similar to the nucleus-nucleus
collisions but the overlap zone is not introduced presently. We deal with
the l+p (l+n) and l+A DIS (SIDIS) like the p+p and p+A collisions,
respectively. However, instead of the $NN$ total cross section, the l+p DIS
total cross section is used and the lepton is assumed not resolvable in the
PYTHIA and PACIAE models.

Fig.~\ref{nccc} shows the leading order (Born approximation) Feynman
diagram for the neutral current (NC, the exchange of $\gamma/Z$ boson,
left panel) and charged current (CC, the exchange of $W^\pm$ boson,
right panel) $e^-$+p DIS. There are two vertices in the left panel of
Fig.~\ref{nccc}, for instance. At the upper boson vertex the initial
state QED and weak radiations have to be considered. At the lower boson
vertex, not only the leading order parton level process of
$V^*q\rightarrow q$ ($V^*$ refers to $\gamma/Z/
W$) but also the first order QCD radiation of $V^*g\rightarrow qg$ as
well as the boson-gluon fusion process of $V^*g\rightarrow q\bar q$ have
to be introduced. Furthermore, the parton shower approach has been
introduced to take higher than first order QCD effects into account
\cite{inge}. Therefore the DIS cross section can be formally expressed
as
\begin{equation}
\sigma_{NC(CC)}=\sigma_{NC(CC)}^{Born}(1+\delta_{NC(CC)}^{qed})
                (1+\delta_{NC(CC)}^{weak})(1+\delta_{NC(CC)}^{qcd})
\end{equation}
\cite{adlo}, where $\sigma_{NC(CC)}^{Born}$ is the Born cross section,
$\delta_{NC(CC)}^{qed}$ and $\delta_{NC(CC)}^{weak}$ are, respectively,
the QED and weak radiative corrections, the QCD radiative correction of
$\delta_{NC(CC)}^{qcd}$ is formally introduced in this paper.

In the lowest-order perturbative QCD theory, the NC/CC DIS Born cross
section of the unpolarized electron on an unpolarized nucleon can be
expressed by the structure functions $F_1, F_2, F_3$ as follows
\cite{pdg}
\begin{equation}\label{epDIS}
\frac{{\rm{d^2}} \sigma _{I}}{{\rm{d}} x {\rm{d}} y}=
\frac{4\pi \alpha^2}{xyQ^2}\eta^I\left( \left(1-y-\frac{x^2y^2M^2}{Q^2}
\right)F_2^I+y^2xF_1^I\mp \left(y-\frac{y^2}{2}\right)xF_3^I\right),
\end{equation}
where the mass of the initial and scattered leptons are neglected. In
the above equation, $I$ denotes $NC$ or $CC$. $\alpha$ stands for the
fine structure constant. $x\equiv x_B$ and $y$ are the Bjorken scaling
variable and fraction energy of $\gamma/Z/W$ boson, respectively. $M$
refers to the mass of target nucleon. $\eta^{NC}=1$, $\eta^{CC}=(1\pm
\lambda)^2\eta_W$, and
\begin{eqnarray}
\eta_W=\frac{1}{2}\left(\frac{G_FM_W^2}{4\pi\alpha}\frac{Q^2}
 {Q^2+M_W^2}\right) \\
G_F=\frac{e^2}{4\sqrt{2}sin^2\theta_WM_W^2},
\end{eqnarray}
where $M_W$ and $\theta_W$ are the mass of $W$ boson and Weinberg angle,
respectively. $\lambda =\pm 1$ is the helicity of the incident lepton.

The structure functions above can be expressed by the parton
distribution function of nucleon in the quark-parton model \cite{bjor}.
Although the PDF can not be calculated by first principle, it can be
extracted from the QCD fits by a measure of the agreement between the
experimental data of lepton-nucleon DIS cross section and the theoretical
models \cite{gizh}. The $e^-$+p DIS total cross section calculated with
HERAPDF1.5 LO \cite{coop} PDF set, is given in Fig.~\ref{tcros} by
black curve \cite{li}. In the calculation the cuts are first set for
$Q^2>1$ GeV and $W^2>1.96$ Gev and then the cuts on $x$ and $y$ are
derived according to relationships of the kinematic variables and
$cos^2\theta\leq 1$. The red and blue circles in Fig.~\ref{tcros} are, respectively, the total cross section of pp and $\gamma$p collisions copied from \cite{pdg}.

One knows well that the incident proton, in the p+Au collisions at RHIC
energies for instance, may collide with a few ($\sim$2-5) nucleons when
it passes through the gold target. Since the $e^-$+p DIS total cross
section is a few order of magnitude small than pp collision at the $\sqrt
s$ range of $\sqrt s<$1000 GeV (cf. Fig.~\ref{tcros}), one may expect
that the incident electron, in this energy range, may suffer at most one
DIS with the nucleon when it passes through the target nucleus. The
struck nucleon is the one with lowest approaching distance from the
incident electron. This is the same for other incident leptons because
the DIS total cross section is not so much different among the different
leptons \cite{li}.

Therefore in the pioneer studies \cite{gisin} for lepton-nucleus DIS by
PYTHIA + BUU transport model, the FRITIOF 7.02 model \cite{pi} or PYTHIA 6.2
\cite{sjos2} was employed to generate lepton-nucleon DIS event. This
generated hadronic final state was then input into the couple channel BUU
(Boltzmann-Uehling-Uhlenbeck) equation \cite{buu} to consider the final state
hadronic interaction (hadronic rescattering). They (Giessen group) employed
this PYTHIA + BUU transport model successfully described the HERMES data of
ratio of the DIS normalized charged hadron multiplicity in the nucleus A to
the one in the deuteron for the 27.5 GeV beam energy $e^++^{14}N$ and
$e^++^{84}Kr$ SIDIS \cite{herm3}. Of course, in the calculations they have
to introduce an assumed values for the lepton-hadron interaction cross
section and the hadronic formation time.

\begin{widetext}
\begin{center}
\begin{figure}[]
\includegraphics[width=0.5\textwidth]{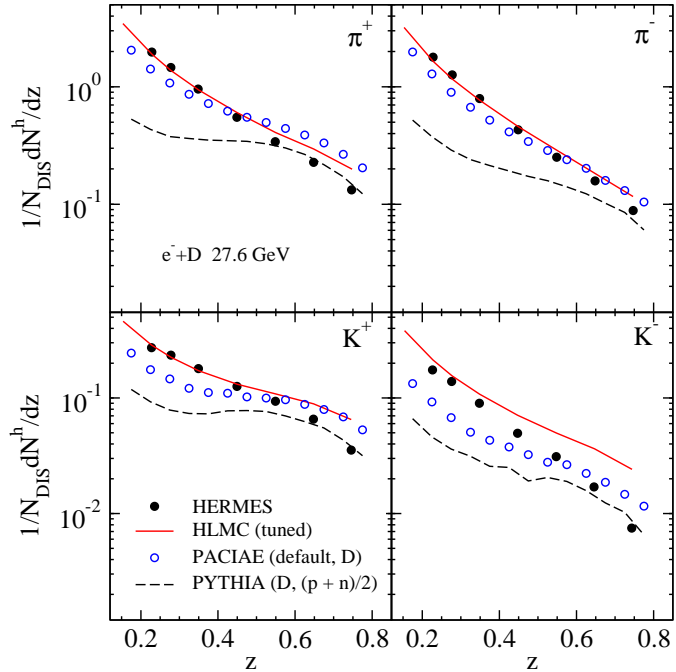}
\caption{(color online) Multiplicity of DIS normalized specific charged
hadron as a function of $z$ in the $e^-+$D SIDIS at 27.6 GeV beam
energy.}
\label{ed45}
\end{figure}
\end{center}
\end{widetext}
\section {Results}
As mentioned in \cite{herm2,hill} the measured hadron multiplicity in
the $e^-$+p and $e^-$+D SIDIS at 27.6 GeV beam energy has first to
correct for the radiative effects, limitations in geometric acceptances,
and the detector resolution. The Born-level multiplicity is then obtained
in order to benefit the PDF extraction, etc. Then they normalized this
Born-level multiplicity by the total DIS yield to reduce the uncertainties
in the corrections above. This is of benefit to the comparison among the
different experimental measurements and between the experiment and theory.
The Born-level multiplicity of the type $h$ hadrons as a function of $z$
in the lepton SIDIS off a nuclear target, for instance, can be expressed
as
\begin{equation}
\frac{1}{N_{DIS}}\frac{dN^h}{dz}=
                  \frac{1}{N_{DIS}}\int d^5N^h(x_B,Q^2,z,P_{h\bot},
                  \phi_h)dx_BdQ^2dP_{h\bot}d\phi_h.
\label{multi}
\end{equation}
Therefore, we can compare the default PYTHIA and PACIAE results of
$\frac{1}{N_{DIS}}\frac{dN^h}{dz}$ calculated in the full kinematic
phase space to the DIS normalized HERMES data, like in \cite{herm2,hill}.

In the default PYTHIA and PACIAE model simulations, the model parameters
are unchanged. The default PYTHIA (black dashed line) and PACIAE (blue
open circles) results of $\frac{1}{N_{DIS}}\frac{dN^h}{dz}$ are compared
with the HERMES data (black solid circles) as well as the results of HLMC
(black line) in the Figure~\ref{ep45} for $\pi^+$ (upper left panel),
$\pi^-$ (upper right), $K^+$ (lower left), and $K^-$ (lower right) in the
$e^-$+p SIDIS at the 27.6 GeV beam energy. One sees in this figure that
the default PACIAE results reproduce HERMES data nearly as good as HLMC
(with thirteen tuned model parameters). However, the default PYTHIA
results disagree with HERMES data. Figure \ref{ed45} is the same as
Fig.~\ref{ep45} but for $e^-$+D SIDIS at the same beam energy. For
Fig.~\ref{ed45} one may draw a similar conclusion like Figure \ref{ep45}.
\begin{widetext}
\begin{center}
\begin{figure}[]
\vspace{1cm}
\includegraphics[width=0.5\textwidth]{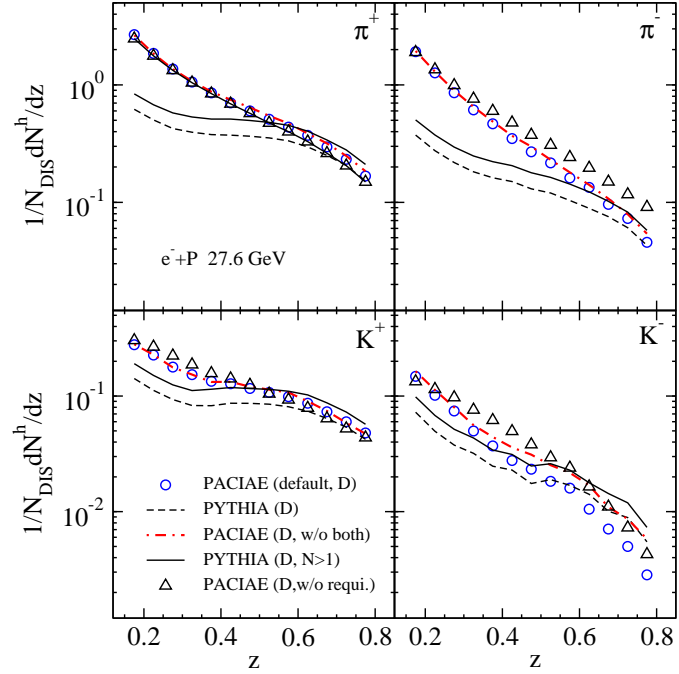}
\caption{(color online) Multiplicity of DIS normalized specific charged
hadron as a function of $z$ in the $e^-+$p SIDIS at 27.6 GeV beam
energy.}
\label{ep45_r}
\end{figure}
\end{center}
\end{widetext}
\begin{widetext}
\begin{center}
\begin{figure}[]
\includegraphics[width=0.5\textwidth]{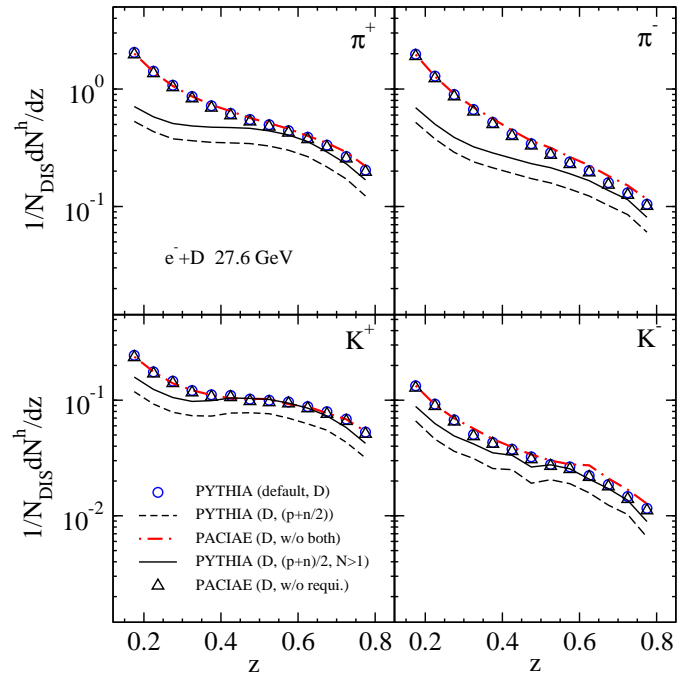}
\caption{(color online) Multiplicity of DIS normalized specific charged
hadron as a function of $z$ in the $e^-+$D SIDIS at 27.6 GeV beam
energy.}
\label{ed45_r}
\end{figure}
\end{center}
\end{widetext}

Table~\ref{diff} lists the discrepancies between the PYTHIA and PACIAE
models. In the event generation there is no extra requirement in the
PYTHIA model but is requirement of having one parton (quark, antiquark,
or gluon) at least in each event in the PACIAE model. In the Figs.
~\ref{ep45_r} and \ref{ed45_r}, the black dashed line is the default
PYTHIA results, the blue circles are the default PACIAE results, and
the red dash-dotted line is the results of PACIAE without both the PRS
and HRS rescatterings. The later two, i.e. blue circles and red
dash-dotted line, are close to each other, which really proves the
small effect of both the PRS and HRS in the $e^-$+p and $e^-$+D SIDIS,
because the reaction systems here are quite small.

The black open triangles, in the figures \ref{ep45_r} and \ref{ed45_r},
are the default PACIAE results calculated without the requirement of
having one parton at least in each generated event. These results are
not different so much from the blue open circles (default PACIAE
results calculated with event requirement). Thus the initial partonic
state, introduced in the PACIAE model but not in PYTHIA, has to be the
main reason of the large discrepancy between the default PYTHIA and
PACIAE results. The introduction of partonic initial state causes
the dynamical simulation processes, such as the partonic rescattering,
the space and time evolutions of hadronization, the hadronic
rescattering, etc., in the PACIAE model are quite different from the
one in the PYTHIA model. The space and time evolutions of hadronization
is especially to be investigated further.

In the Figures \ref{ep45_r} and \ref{ed45_r} the black solid lines
are calculated by the default PYTHIA model with the requirement of
having one pion or kaon at least in each generated event. One sees
here that the default PYTHIA results with the requirement (black
solid line) is considerably larger than the one without the
requirement (black dashed line). It may mean that the event generated
by the default PYTHIA is not completely DIS, and may confuse with
diffractive processes etc., which has to be study further.

\begin{table*}[htbp]
\tabcolsep 2.mm
\caption{Discrepancies between the PYTHIA and PACIAE models}
\begin{tabular}{ccc}
\hline \hline
item &PYTHIA &PACIAE\\
\hline
Partonic initial state &not introduced &introduced \\
Initial state PRS &no &yes \\
Final state HRS &no &yes \\
Event requirement &no &yes \\
\hline \hline
\end{tabular}
\label{diff}
\end{table*}
\section {Conclusions}
In summary, we have employed the PYTHIA 6.4 and the extended parton
and hadron cascade model PACIAE 2.2 to investigate the DIS normalized
specific charged hadron multiplicity, $\frac{1}{N_{DIS}}\frac{dN^h}{dz}$,
measured by HERMES in the $e^-$+P and $e^-$+D SIDIS at 27.6 GeV bean
energy. The PYTHIA and PACIAE results, calculated with default model
parameters, not well and fairly well reproduce the HERMES data
\cite{herm2}, respectively.

Additionally, we have investigated the effects of the differences between
the PYTHIA and PACIAE models, i.e the effect of both the initial state
PRS and final state HRS as well as the event requirement. It turned out
that the effect of both the PRS and HRS is weak because of the small
reaction system of $e^-$+p and $e^-$+D SIDIS. The event requirement of
having one parton (quark, antiquark, or gluon) at least in each generated
initial partonic state introduced in the PACIAE model plays a visible
role. However the effect of the requirement of having at least one pion
or kaon in each event generated by the PYTHIA model is relatively strong.
The main reason, which causes a discrepancy between the default PACIAE
and PYTHIA results, should be attributed to the initial partonic state,
introduced in the PACIAE model but not in PYTHIA. This discrepancy leads
to the simulation processes, such as the partonic rescattering, the space
and time evolutions of hadronization, the hadronic rescattering, etc.,
in the PACIAE model are quite different from the PYTHIA model. The more
investigations are especially required for the space and time evolutions
of hadronization.

Acknowledgments: This work was supported by the National Natural Science
Foundation of China under grant Nos.:11175070, 11205066, 11175262
11477130, 11375069 and by the 111 project of the foreign expert bureau of China. BHS would like to thank C. P. Yuan for HERAFitter, G. Schnell for HERMES data, and Y. Mao and S. Joosten for discussions. YLY acknowledges the
financial support from SUT-NRU project under contract No. 17/2555.

\end{document}